\documentclass{article} 
\usepackage{iclr2025_conference,times}

\usepackage{amsmath,amsfonts,bm}

\def\eqref#1{equation~\ref{#1}}

\def\1{\bm{1}}

\DeclareMathAlphabet{\mathsfit}{\encodingdefault}{\sfdefault}{m}{sl}
\SetMathAlphabet{\mathsfit}{bold}{\encodingdefault}{\sfdefault}{bx}{n}

\usepackage{hyperref}
\usepackage{url}
\usepackage{booktabs}
\usepackage{graphicx}
\usepackage{algorithm}
\usepackage{algorithmicx}
\usepackage{algpseudocode}

\title{Synthetic Poisoning Attacks: The Impact of Poisoned MRI Image on U-Net Brain Tumor Segmentation}

\author{Tianhao Li$^{1}$\thanks{Equal contribution} \quad Tianyu Zeng$^{2}$\footnotemark[1] \quad Yujia Zheng$^{1,3,4}$\footnotemark[1]  \quad Chulong Zhang$^{1}$ \quad Jingyu Lu$^{5}$ \\  \textbf{Haotian Huang}$^{3}$ \quad
\textbf{Chuangxin Chu}$^{6}$ \quad  \textbf{Fang-Fang Yin}$^{7}$  \quad \textbf{Zhenyu Yang}$^{7}$ 
\AND\vspace{-2em}
\\
$^{1}$Duke University, 
$^{2}$Hong Kong Polytechnic University, 
$^{3}$North China University of Technology,\\
$^{4}$State Key Laboratory of Intelligent Game, Institute of Software Chinese Academy of Sciences, \\
$^{5}$Australian National University,
$^{6}$Nanyang Technological University, 
$^{7}$Duke Kunshan University.\\
\\
\small{
 \texttt{
   \{\href{mailto:tianhao.li2@duke.edu}{tianhao.li2}, \href{mailto:zy84@duke.edu}{zy84}\}@duke.edu
 }}
}

\iclrfinalcopy 
\begin{document}

\maketitle

\begin{abstract}
Deep learning-based medical image segmentation models, such as U-Net, rely on high-quality annotated datasets to achieve accurate predictions. However, the increasing use of generative models for synthetic data augmentation introduces potential risks, particularly in the absence of rigorous quality control. In this paper, we investigate the impact of synthetic MRI data on the robustness and segmentation accuracy of U-Net models for brain tumor segmentation. Specifically, we generate synthetic T1-contrast-enhanced (T1-Ce) MRI scans using a GAN-based model with a shared encoding-decoding framework and shortest-path regularization. To quantify the effect of synthetic data contamination, we train U-Net models on progressively ``poisoned'' datasets, where synthetic data proportions range from 16.67\% to 83.33\%. Experimental results on a real MRI validation set reveal a significant performance degradation as synthetic data increases, with Dice coefficients dropping from 0.8937 (33.33\% synthetic) to 0.7474 (83.33\% synthetic). Accuracy and sensitivity exhibit similar downward trends, demonstrating the detrimental effect of synthetic data on segmentation robustness. These findings underscore the importance of quality control in synthetic data integration and highlight the risks of unregulated synthetic augmentation in medical image analysis. Our study provides critical insights for the development of more reliable and trustworthy AI-driven medical imaging systems.
\end{abstract}

\section{Introduction}

Deep learning-based segmentation models \citep{minaee2021image}, such as U-Net \citep{ronneberger2015u}, have demonstrated remarkable success in medical image analysis \citep{azad2024medical, du2020medical}, particularly in brain tumor segmentation tasks \citep{abidin2024recent, ranjbarzadeh2023brain, das2022artificial, magadza2021deep}. These models rely heavily on high-quality image-segmentation pairs to ensure accurate and reliable predictions. However, the growing adoption of generative models for synthetic medical image creation introduces new challenges \citep{dayarathna2024deep}. While synthetic data can potentially augment training datasets, improve data diversity, and address class imbalances, its uncontrolled incorporation may lead to significant performance degradation \citep{hao2024synthetic}. Without rigorous quality control, synthetic data can act as a form of ``data poisoning'', negatively impacting model robustness and segmentation accuracy \citep{yerlikaya2022data, pitropakis2019taxonomy}.

In recent years, generative adversarial networks (GANs) \citep{goodfellow2014generative, goodfellow2020generative} have emerged as a popular technique for generating synthetic medical images \citep{alamir2022role, singh2021medical, nie2017medical}. These models leverage learned distributions from real data to synthesize realistic samples. While some studies have explored the benefits of GAN-generated data for augmentation \citep{makhlouf2023use, zhang2023gan, chen2022generative, hatamizadeh2021swin}, few have systematically examined the risks associated with using synthetic medical images in segmentation tasks. Specifically, the effects of synthetic data contamination on segmentation models remain insufficiently studied, raising concerns about potential accuracy degradation and unreliable clinical applications \citep{singkorapoom2023pre}.

To address such a gap, we evaluate the impact of synthetic MRI data on the performance of U-Net \citep{ronneberger2015u} models for brain tumor segmentation. We consider synthetic data as a type of contamination and investigate how increasing proportions of synthetic data influence segmentation robustness. Using a novel GAN-based model \citep{xie2023unpaired}, we generate synthetic T1-contrast-enhanced (T1-Ce) MRI images from paired CT-MRI datasets and introduce them into training sets at varying proportions. 
We compare a baseline U-Net trained exclusively on real MRI data against U-Net models trained on progressively “poisoned” datasets containing increasing amounts of synthetic data. Segmentation performance is assessed using Dice coefficient, Jaccard index, accuracy, and sensitivity to quantify the extent of degradation. The results demonstrate a significant decline in segmentation performance as the proportion of synthetic data increases, with notable drops in Dice coefficients, Jaccard index, and sensitivity. These findings emphasize the importance of establishing best practices for incorporating synthetic data in medical image segmentation pipelines. 
By highlighting the potential risks of synthetic data contamination, this study provides valuable insights for developing robust and trustworthy deep learning \citep{wang2023trustworthy, li2023trustworthy, huang2018safety, hanif2018robust, li2024p, zheng20245} in medical imaging applications \citep{shukla2023trustworthy, teng2024literature, fidon2023trustworthy, shi2024survey}.

\vspace{1em}

\section{Related Works}

\paragraph{Brain Tumor Segmentation}  
Brain tumor segmentation has been a critical task in medical image analysis, enabling precise delineation of tumor regions for diagnosis and treatment planning \citep{wadhwa2019review, gordillo2013state}. Traditional methods relied on handcrafted features \citep{mecheter2022deep, khan2020cascading, hasan2019combining} and classical machine learning models \citep{soomro2022image, amin2019brain, bakas2018identifying}, but deep learning approaches, particularly convolutional neural networks (CNNs) \citep{li2021survey}, have significantly advanced segmentation performance \citep{havaei2017brain, pereira2016brain}. U-Net \citep{ronneberger2015u} and its variants \citep{azad2024medical, siddique2021u} have become the backbone of many segmentation pipelines due to their encoder-decoder architecture and skip connections, which preserve spatial information. More recent methods leverage transformer-based architectures \citep{ghazouani2024efficient, wang2023vision, jiang2022swinbts, huang2022transformer} and hybrid CNN-Transformer models \citep{liu2024transsea, kang2024multimodal, chen2022csu, jia2021bitr} to enhance feature representation and improve segmentation accuracy. Despite these advancements, the robustness of segmentation models remains a concern, especially when trained on datasets with varying levels of synthetic content.

\paragraph{GAN-based MRI Synthesis}  
The generation of synthetic MRI images has gained significant attention due to its potential to augment datasets, address data scarcity, and enable cross-modality learning \citep{tiwari2025review, choi2025beyond, pani2024synthetic, koetzier2024generating, hamghalam2024medical, ji2022synthetic, han2018gan, blystad2012synthetic}. Generative adversarial networks (GANs) \citep{goodfellow2014generative, goodfellow2020generative} and variational autoencoders (VAEs) \citep{kingma2013auto, pinheiro2021variational} have been widely explored for MRI synthesis \citep{tavse2022systematic, laptev2021generative}. Conditional GANs \citep{mirza2014conditional} and cycle-consistent GANs \citep{zhu2017unpaired} have been applied for modality translation, such as generating MRI from CT scans. More recent works incorporate structural constraints and perceptual losses to improve anatomical consistency in synthetic images. However, concerns persist regarding the quality and fidelity of synthetic images, as even minor artifacts or inconsistencies can propagate through downstream tasks, adversely affecting segmentation performance. In this context, synthetic medical images may act as a form of data poisoning, compromising model reliability and clinical applicability \citep{singkorapoom2023pre}.

\paragraph{Data Poisoning Attack}  
Data poisoning attacks involve injecting manipulated, low-quality, or malicious data into training datasets to degrade model performance or induce adversarial vulnerabilities \citep{yerlikaya2022data, fan2022survey}. In the medical imaging domain, poisoning can occur through mislabeled \citep{tolpegin2020data, lin2021active}, perturbed \citep{martinelli2023data, bortsova2021adversarial}. Prior research has demonstrated that even small perturbations in training data can lead to significant performance degradation in classification and segmentation models \citep{szegedy2013intriguing, chakraborty2021survey}. While poisoning attacks and corresponding mitigation strategies \citep{goldblum2022dataset, schwarzschild2021just, fu2024poisonbench, li2024scisafeeval} have been extensively studied in general computer vision tasks \citep{wei2024physical, akhtar2018threat}, their impact on medical imaging pipelines remains underexplored. Our work investigates the effect of synthetic MRI data as a form of data poisoning, evaluating its impact on brain tumor segmentation performance.

\section{Methods}

\begin{figure}[htbp]
    \centering
    \includegraphics[width=\linewidth]{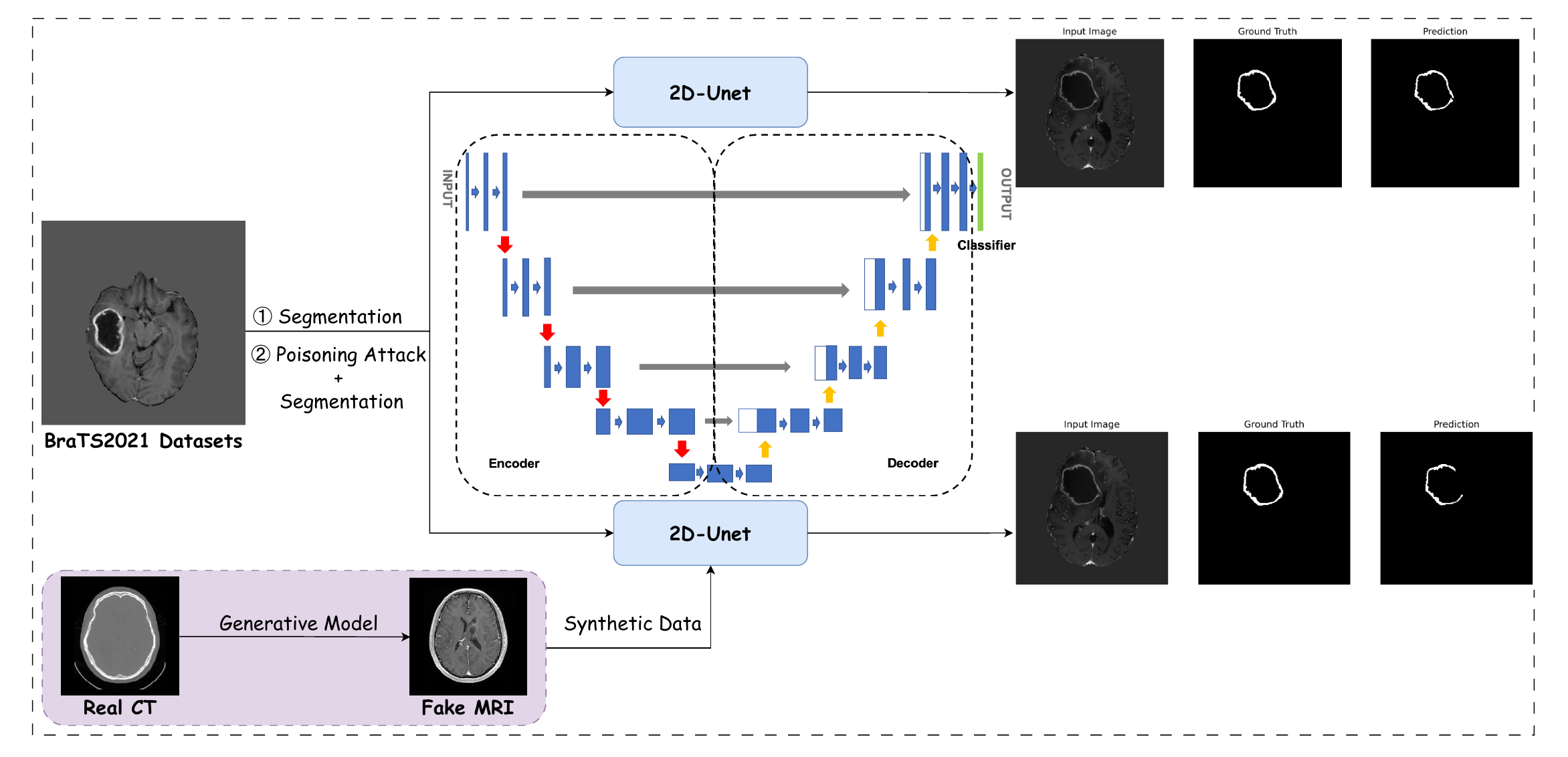}
    \vspace{-2em}
    \caption{Overall workflow.}
    \label{fig:overall}
\end{figure}

In this study, we investigate the potential degradation in segmentation accuracy when synthetic data is incorporated into the training process, an overall workflow is shown in Figure \ref{fig:overall}. Formally, let $M$ denote a medical AI model trained for segmentation tasks, and let $S$ represent an image synthesis model designed to generate synthetic medical images. We define a real medical dataset as $D$, and a modified dataset containing synthetic samples generated using $S$ as $D'$. Our objective is to analyze the effects of training $M$ on $D'$.

\begin{figure}[t]
    \centering
    \includegraphics[width=\linewidth]{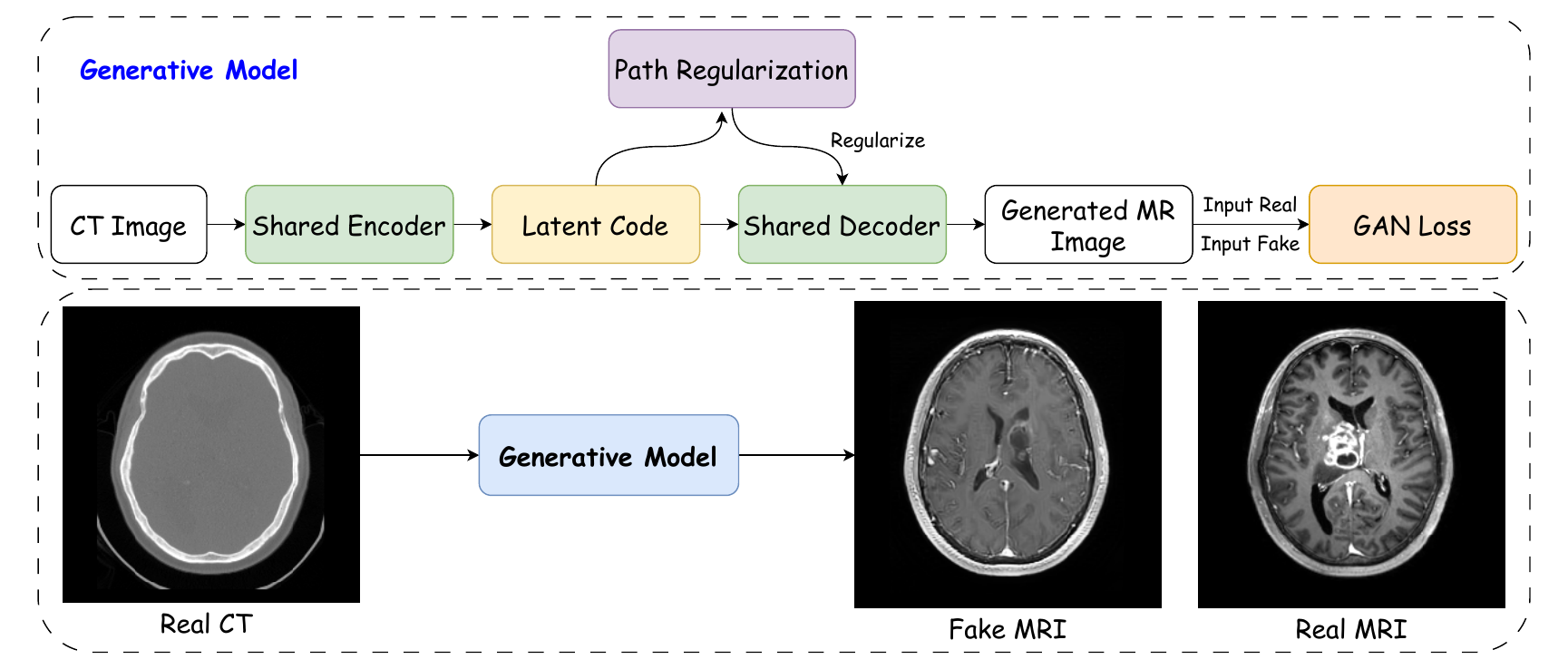}
    \vspace{-2em}
    \caption{Workflow of generative model \citep{xie2023unpaired}.}
    \label{fig:workflow}
\end{figure}

Specifically, as shown in Algorithm \ref{alg:synthetic_contamination}, our study follows these steps: (\romannumeral1) We first prepare a medical dataset $D$, and employ a generative model $S$ \citep{xie2023unpaired}, as shown in Figure \ref{fig:workflow}, to produce synthetic data, resulting in a modified dataset $D'$. (\romannumeral2) We then train a baseline model $M$ trained solely on $D$, and a model $M'$ trained on $D'$, which contains synthetic images. (\romannumeral3) We evaluate the segmentation performance of both models $M$ and $M'$ using metrics including the Dice coefficient, Jaccard index, accuracy, and sensitivity. 

\begin{algorithm}[htbp]
\caption{Training and Evaluation of U-Net with Synthetic Data}
\label{alg:synthetic_contamination}
\begin{algorithmic}[1]
\Require $\mathcal{D} = \{(x, y)\}$, $S$, $\mathcal{P} = \{16.67\%, 33.33\%, 50.00\%, 66.67\%, 83.33\%\}$

\For{$p \in \mathcal{P}$} \Comment{Poisoning}
    \State $\mathcal{X'} \gets S(\mathcal{D})$
    \State $\mathcal{D'}(p) \gets \mathcal{D} \cup \mathcal{X'}$
\EndFor

\State $\mathcal{M} \gets$ Train U-Net on $\mathcal{D}$
\For{$p \in \mathcal{P}$} \Comment{Training}
    \State $\mathcal{M'}(p) \gets$ Train U-Net on $\mathcal{D'}(p)$
\EndFor

\For{$p \in \mathcal{P}$} \Comment{Evaluation}
    \State Compute $\text{Dice}(\mathcal{M}), \text{Dice}(\mathcal{M'}(p))$
    \State Compute Jaccard, Accuracy, Sensitivity
\EndFor

\end{algorithmic}
\end{algorithm}

We expect that the segmentation performance of $M'$ will be lower than that of $M$, formally expressed as: $\text{Dice}(M') < \text{Dice}(M)$, where $\text{Dice}(M)$ and $\text{Dice}(M')$ represent the Dice coefficients of the models trained on $D$ and $D'$, respectively. Through systematic experimentation, we aim to quantify the extent of performance degradation and provide insights into the risks associated with synthetic data contamination in medical image segmentation tasks.

\section{Experiments}

\subsection{Setup}

\paragraph{Dataset}
We conduct our experiments using the publicly available BraTS2021 dataset, which contains T1-contrast-enhanced (T1-Ce) MRI scans from 150 glioma patients along with their corresponding enhanced tumor (ET) segmentation masks \citep{menze2014multimodal}. To introduce synthetic data, we employ a generative adversarial network (GAN) \citep{goodfellow2020generative, goodfellow2014generative} designed for cross-domain medical image translation \citep{xie2023unpaired}. Specifically, the GAN is trained on 660 paired CT-MRI datasets and features a shared encoding-decoding framework with shortest-path regularization \citep{xie2023unpaired} to ensure anatomical consistency during translation. Some cases of synthetic MRI can be found in the appendix \ref{sec:fake_mri}. The trained model generates 150 synthetic T1-Ce MRI scans, which are then incorporated into our training set at varying proportions.

\paragraph{Protocol}
We evaluate the impact of synthetic MRI data on U-Net segmentation performance. Let $\mathcal{D} = \{(x, y)\}$ represent the original dataset, where $x$ denotes real MRI scans and $y$ the corresponding segmentation masks. Using the trained GAN model, we generate synthetic MRI images $\mathcal{X'}$. We construct modified datasets $\mathcal{D'}(p)$ by mixing real MRI scans with synthetic samples, where $p \in \{16.67\%, 33.33\%, 50.00\%, 66.67\%, 83.33\%\}$ represents the proportion of synthetic data. Two types of U-Net models are trained:  
(\romannumeral1) A baseline model $\mathcal{M}$ trained solely on $\mathcal{D}$, establishing a performance reference.  
(\romannumeral2) Poisoned models $\mathcal{M'}(p)$ trained on $\mathcal{D'}(p)$, simulating different levels of synthetic data contamination.  

\paragraph{Metrics}
We evaluate the segmentation performance of $\mathcal{M}$ and $\mathcal{M'}(p)$ on a real MRI test set using the following standard metrics:  

(\romannumeral1) \textit{Dice Coefficient}: Measures the spatial overlap between the predicted segmentation $\hat{Y}$ and the ground truth $Y$. Defined as:  
\begin{equation}
\text{Dice} = \frac{2 |\hat{Y} \cap Y|}{|\hat{Y}| + |Y|}
\end{equation}
where $|\hat{Y} \cap Y|$ represents the number of correctly segmented pixels, and $|\hat{Y}|$ and $|Y|$ denote the total number of pixels in the predicted and ground truth masks, respectively. A higher Dice score indicates better segmentation performance.  

(\romannumeral2) \textit{Jaccard Index}: Also known as the Intersection-over-Union (IoU), this metric evaluates the proportion of correctly segmented pixels relative to the union of predicted and ground truth segmentations:  
\begin{equation}
\text{Jaccard} = \frac{|\hat{Y} \cap Y|}{|\hat{Y} \cup Y|}
\end{equation}
Jaccard provides a stricter evaluation compared to Dice, as it penalizes false positives and false negatives more severely.  

(\romannumeral3) \textit{Accuracy}: Measures the overall correctness of pixel classification, considering both the segmented tumor region and the background:  
\begin{equation}
\text{Accuracy} = \frac{TP + TN}{TP + TN + FP + FN}
\end{equation}
where $TP$, $TN$, $FP$, and $FN$ represent true positives, true negatives, false positives, and false negatives, respectively. Accuracy alone can be misleading in imbalanced segmentation tasks, where background pixels dominate.  

(\romannumeral4) \textit{Sensitivity}: Also known as recall or true positive rate, sensitivity quantifies the model’s ability to correctly identify tumor regions:  
\begin{equation}
\text{Sensitivity} = \frac{TP}{TP + FN}
\end{equation}
A higher sensitivity indicates fewer missed tumor regions, which is critical for medical applications where under-segmentation could lead to misdiagnoses.

\subsection{Results}

\begin{figure}[t]
    \centering
    \includegraphics[width=\linewidth]{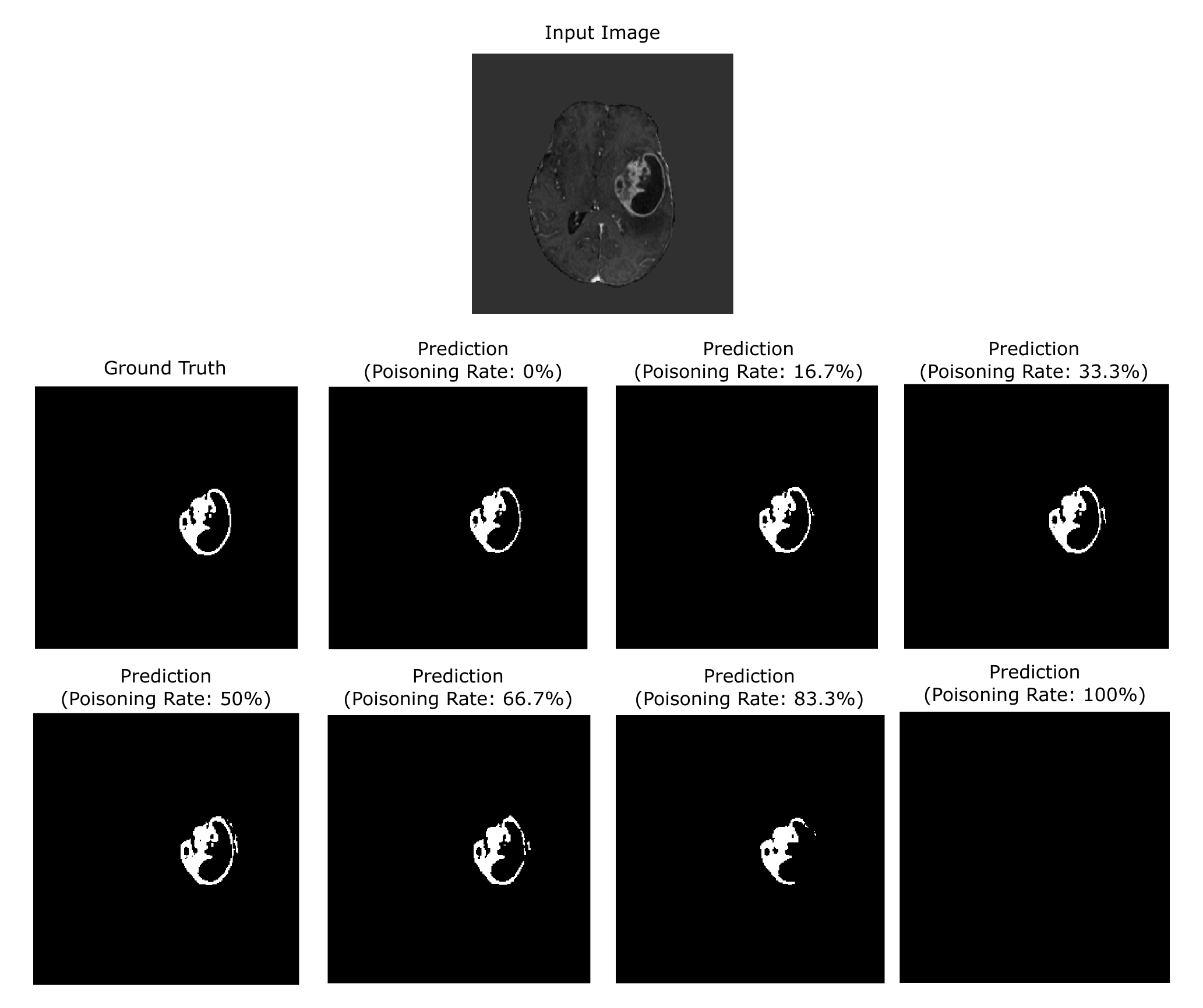}
    \vspace{-2em}
    \caption{A sample of segmentation results of the ET region from the same MRI scan using the U-Net model at varying poisoning rates.}
    \label{fig:sample_results}
\end{figure}

Figure \ref{fig:sample_results} illustrates the segmentation outputs of the U-Net model under varying poisoning rates. At low synthetic data proportions (e.g., 16.67\%), model predictions remain close to the ground truth. However, as $p$ increases, segmentation quality deteriorates, with higher poisoning levels leading to incorrect tumor boundary delineations. Table \ref{tab:result} presents the quantitative impact of synthetic data contamination. The Dice coefficients decrease from 0.8937 ($p=33.33\%$) to 0.7474 ($p=83.33\%$), confirming a strong correlation between synthetic data proportion and segmentation degradation. Jaccard and sensitivity exhibit similar trends, with significant performance drops beyond $p=50\%$. However, the increase of a portion of the synthetic data has a negligible effect on accuracy.
These findings suggest that, while low proportions of synthetic data may not drastically harm model performance, excessive reliance on synthetic data compromises segmentation robustness.  

\begin{table}[t]
\caption{Quantitative description of metrics on varying poisoning rates}
\begin{tabular}{@{}ccccc@{}}
\toprule
\textbf{Poisoning Rate(\%)} & \textbf{Dice}   & \textbf{Jaccard} & \textbf{Accuracy} & \textbf{Sensitivity} \\ \midrule
0.00                        & 0.8939 ± 0.1243 & 0.8184 ± 0.1546  & 0.9983 ± 0.0011   & 0.9136 ± 0.1578      \\
16.67                       & 0.8650 ± 0.2072 & 0.7937 ± 0.2151  & 0.9638 ± 0.1854   & 0.8790 ± 0.2390      \\
33.33                       & 0.8937 ± 0.0722 & 0.8145 ± 0.1071  & 0.9981 ± 0.0013   & 0.9191 ± 0.1173      \\
50.00                       & 0.8572 ± 0.1580 & 0.7738 ± 0.1810  & 0.9979 ± 0.0011   & 0.9292 ± 0.1208      \\
66.67                       & 0.8146 ± 0.2457 & 0.7360 ± 0.2458  & 0.9978 ± 0.0013   & 0.8328 ± 0.2817      \\
83.33                       & 0.7474 ± 0.2650 & 0.6486 ± 0.2579  & 0.9967 ± 0.0020   & 0.7577 ± 0.3054      \\ \bottomrule
\end{tabular}
\label{tab:result}
\end{table}

To further understand the effects of synthetic MRI data, we analyze segmentation performance across different poisoning thresholds. Our results indicate that models trained with $p\leq 33.33\%$ maintain relatively stable performance, while those with $p\geq 50\%$ suffer from severe degradation. This highlights the importance of synthetic data curation, suggesting that controlled synthetic augmentation may be feasible if appropriately regulated.  

Our findings raise critical concerns about the integration of synthetic medical images in training pipelines. While synthetic MRI augmentation can be beneficial in low proportions, excessive synthetic data exposure introduces model biases and reduces segmentation reliability. These results emphasize the need for quality control mechanisms and hybrid training strategies that combine real and synthetic data optimally to mitigate potential risks in medical AI applications.

\section{Conclusion}

We investigated the impact of synthetic MRI data on the robustness and segmentation accuracy of U-Net models for brain tumor segmentation. Experiment results suggest that the inclusion of synthetic data, when not properly regulated, significantly degrades segmentation performance. As the proportion of synthetic MRI data increased, we observed a substantial decline in key evaluation metrics, including Dice coefficient, Jaccard index, accuracy, and sensitivity. Our findings highlight that while small proportions of synthetic data may not drastically impair model performance, excessive reliance on synthetic samples introduces severe biases, compromises segmentation reliability, and leads to inaccurate tumor boundary delineations. We provide crucial insights for designing safer, more reliable deep learning models in medical imaging. As the adoption of generative models continues to expand, our work serves as a foundation for establishing best practices in the responsible integration of synthetic data in AI-driven healthcare systems.

\bibliography{iclr2025_conference}
\bibliographystyle{iclr2025_conference}

\newpage
\appendix
\section{Synthetic MRI Results} \label{sec:fake_mri}

\begin{figure}[h]
    \centering
    \includegraphics[width=0.9\linewidth]{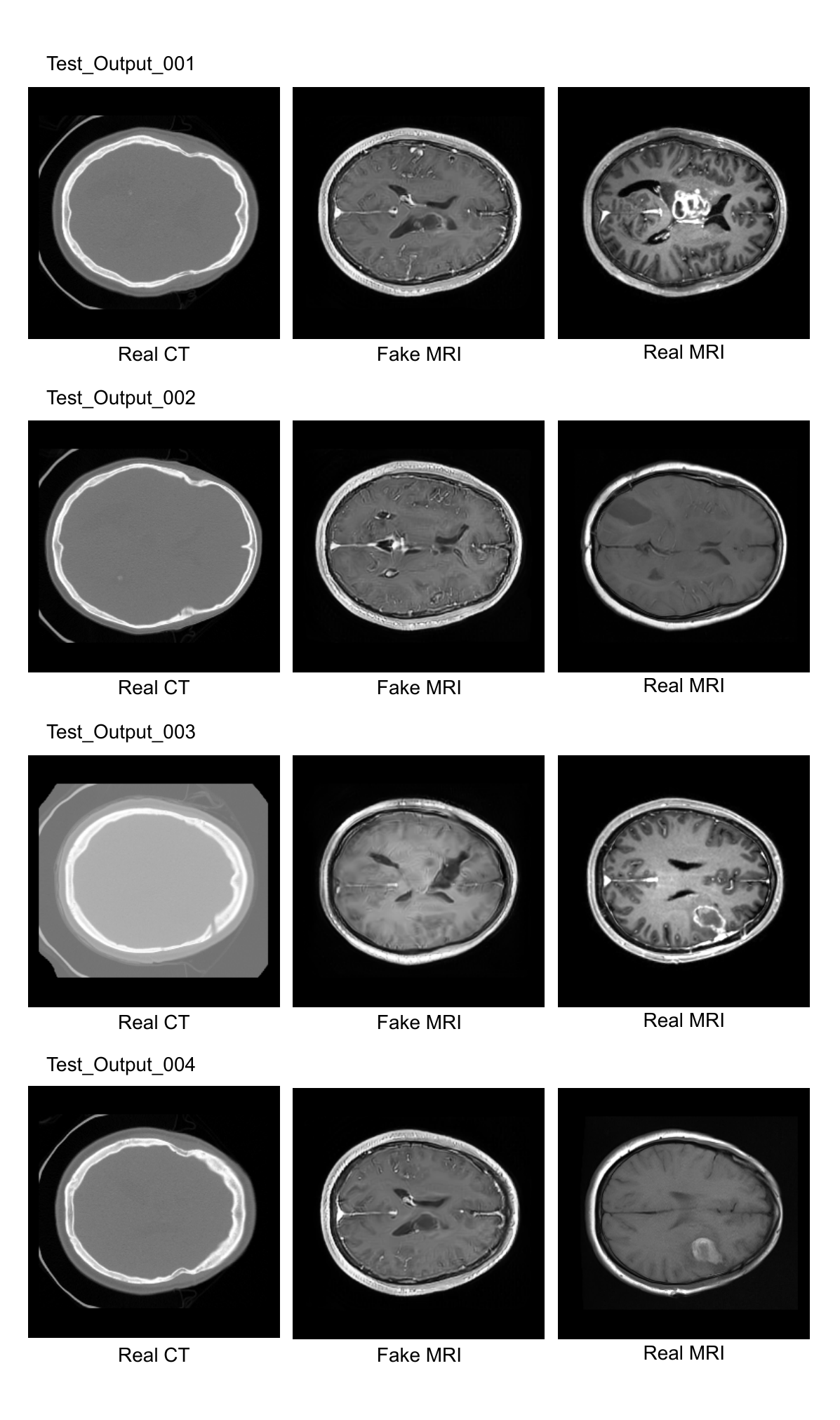}
    \vspace{-3em}
    \caption{Case studies of synthetic MRI from real CT.}
    \label{fig:fake_mri}
\end{figure}

We use a generative model \citep{xie2023unpaired} to synthesize fake MRI from real CT. Figure \ref{fig:fake_mri} presents comparative case studies of fake MRI and real MRI.
\vspace{-1em}

\end{document}